\documentclass[sigconf]{acmart}

\AtBeginDocument{%
  \providecommand\BibTeX{{%
    \normalfont B\kern-0.5em{\scshape i\kern-0.25em b}\kern-0.8em\TeX}}}

\setcopyright{acmcopyright}
\copyrightyear{2018}
\acmYear{2018}
\acmDOI{XXXXXXX.XXXXXXX}

\usepackage{soul}
\usepackage{xcolor}
\usepackage{xspace}
\usepackage{url}
\usepackage{balance}
\usepackage{graphicx}
\usepackage{listings,multicol,multirow}
\usepackage[caption=false]{subfig}  
\usepackage[export]{adjustbox}
\usepackage{amsmath}
\usepackage{textcomp}
\usepackage{breakurl} 
\usepackage{cleveref}
\usepackage{algorithm}
\usepackage[noend]{algpseudocode}
\usepackage{diagbox}
\usepackage{fixfoot}
\usepackage{enumerate}
\usepackage[shortlabels]{enumitem}
\usepackage{tikz}
\newcommand*\circled[1]{\tikz[baseline=(char.base)]{
            \node[shape=circle,draw,inner sep=2pt] (char) {#1};}}

\newcommand{\javasubmissionsinitial}{1,065,579\xspace\xspace}
\newcommand{\javastudents}{37,626\xspace}
\newcommand{\javasteps}{415\xspace}

\newcommand{\validationsteps}{85\xspace}

\copyrightyear{2023}
\acmYear{2023}
\setcopyright{acmlicensed}\acmConference[ITiCSE 2023]{Proceedings of the 2023 Conference on Innovation and Technology in Computer Science Education V. 1}{July 8--12, 2023}{Turku, Finland}
\acmBooktitle{Proceedings of the 2023 Conference on Innovation and Technology in Computer Science Education V. 1 (ITiCSE 2023), July 8--12, 2023, Turku, Finland}
\acmPrice{15.00}
\acmDOI{10.1145/3587102.3588800}
\acmISBN{979-8-4007-0138-2/23/07}

\begin{document}

\title[Detecting Code Quality Issues in Pre-written Templates of Programming Tasks in Online Courses]{Detecting Code Quality Issues in Pre-written Templates \\ of Programming Tasks in Online Courses}


\author{Anastasiia Birillo}
\affiliation{
  \institution{JetBrains Research}
  \city{Belgrade}
  \country{Republic of Serbia}
}
\email{anastasia.birillo@jetbrains.com}

\author{Elizaveta Artser}
\affiliation{
  \institution{Constructor University}
  \city{Bremen}
  \country{Germany}
}
\email{eartser@constructor.university}

\author{Yaroslav Golubev}
\affiliation{
  \institution{JetBrains Research}
  \city{Belgrade}
  \country{Republic of Serbia}
}
\email{yaroslav.golubev@jetbrains.com}

\author{Maria Tigina}
\affiliation{
  \institution{JetBrains Research}
  \city{Belgrade}
  \country{Republic of Serbia}
}
\email{maria.tigina@jetbrains.com}

\author{Hieke Keuning}
\affiliation{
  \institution{Utrecht University}
  \city{Utrecht}
  \country{The Netherlands}
}
\email{h.w.keuning@uu.nl}

\author{Nikolay Vyahhi}
\affiliation{
  \institution{Stepik}
  \city{Boston, MA}
  \country{USA}
}
\email{vyahhi@stepik.org}

\author{Timofey Bryksin}
\affiliation{
  \institution{JetBrains Research}
  \city{Limassol}
  \country{Republic of Cyprus}
}
\email{timofey.bryksin@jetbrains.com}

\begin{abstract}
  In this work, we developed an algorithm for detecting code quality issues in the templates of online programming tasks, validated it, and conducted an empirical study on the dataset of student solutions. The algorithm consists of analyzing recurring unfixed issues in solutions of different students, matching them with the code of the template, and then filtering the results. Our manual validation on a subset of tasks demonstrated a precision of 80.8\% and a recall of 73.3\%. We used the algorithm on 415 Java tasks from the JetBrains Academy platform and discovered that as much as 14.7\% of tasks have at least one issue in their template, thus making it harder for students to learn good code quality practices. We describe our results in detail, provide several motivating examples and specific cases, and share the feedback of the developers of the platform, who fixed 51 issues based on the output of our approach. 
\end{abstract}


\maketitle

\section{Introduction}

Online learning gains popularity and reaches more and more areas~\cite{boehm2006view}, such as foreign languages, economics, natural sciences, and, of course, computer science and programming.
Along with traditional courses in schools and universities~\cite{brusilovsky2014increasing}, many \textit{massive open online courses (MOOCs)} emerge for learning programming on different educational platforms~\cite{oh2020design}.
A recent trend in such platforms is to not only check the correctness of tasks but also pay attention to the quality of the code~\cite{keuning2017code, borstler2018know, keuning2019teachers}.

Nowadays, there exist many novice-friendly tools for checking code quality~\cite{blau2015frenchpress, ureel2019automated, choudhury2016scale, keuning2021tutoring, birillo2022hyperstyle}. 
These tools are used both within classes at universities and on large educational platforms such as JetBrains Academy~\cite{hyperskill}, Leetcode~\cite{leetcode}, or Khan Academy~\cite{khanacademy}.
Many studies show the effectiveness of using such tools, with students' code becoming of higher quality~\cite{keuning2020student, de2018understanding, edwards2017investigating, techapalokul2017understanding, bai2019amelioration, birillo2022hyperstyle}. 
In addition, there is a lot of research aimed at identifying the most prevalent code quality issues in student code or identifying the reasons why these issues may not be corrected~\cite{keuning2017code, de2018understanding, edwards2017investigating, albluwi2020using, aivaloglou2016kids, techapalokul2017understanding}.
However, these results are based on the assumption that all issues in the code were made by the students themselves.

The key problem here is that many tasks, especially on large educational platforms, contain pre-written code (code \textit{templates}) that students complete with their solution.
Such code templates may initially contain code quality issues, but the code quality will be checked only after submitting the correct solution. 
Moreover, to identify issues in templates, it is not enough to run some tool to detect problems on the template itself, since false-positive cases may occur.
For example, a template might define a variable that is intended to be used in the solution. However, in the template itself, this variable will be considered unused, triggering a corresponding warning of code quality tools. 
To the best of our knowledge, there are no studies examining code quality issues in code templates. 

In this pilot study of code quality issues in templates, we explore different tasks in Java on the JetBrains Academy platform~\cite{hyperskill}, which uses the \textit{Hyperstyle} tool~\cite{birillo2022hyperstyle} to check the quality of students' code and report the detected issues.
Firstly, we developed an algorithm to detect code quality issues in task templates.
The algorithm can handle issues from any code quality assessment tool, since it works only with its output.
Next, we collected a dataset of \javasteps Java tasks and \javasubmissionsinitial student submissions over a one-year period.
We manually validated the algorithm on 85 tasks (about 20\,\% of the dataset), with the approach demonstrating a precision of 80.8\,\% and a recall of 73.3\,\%.
Finally, we answered two research questions based on the data in the collected dataset: (1) How often do templates contain code quality issues, and (2) What are the most common code quality issues in templates. 
The results showed that as many as 14.7\,\% of tasks in our dataset had at least one code quality issue in their templates, and that there were specific tasks with an abnormally high number of issues.
These results were reported to the developers of the JetBrains Academy platform, who positively commented on their usefulness, fixed 51 of the discovered issues in 27 tasks, and plan to fix the rest later.

To sum up, with this work we make the following contributions:
\begin{itemize}
    \item An \textbf{algorithm} that detects code quality issues in pre-written templates of programming tasks with a precision of 80.8\,\% and a recall of 73.3\,\%. The source code is available online~\cite{artifacts}.
    \item An \textbf{analysis} of how often code quality issues appear in the templates of 415 Java tasks from the JetBrains Academy platform, and which issues are the most prevalent.
    \item \textbf{Insights} and discussion into the nature of these issues, including important aspects of detecting them, specific examples, and potential origins.
\end{itemize}
\section{Related work and Background}
\label{sec:RW}

\textbf{Code quality in education}. Code quality issues that students introduce when learning to program are currently being actively studied.
Code quality studies come in many forms --- some studies focus on developing tools to find code quality issues~\cite{choudhury2016scale, ureel2019automated, keuning2021tutoring, birillo2022hyperstyle}, while some research student submissions and code quality issues within them~\cite{keuning2017code, de2018understanding, edwards2017investigating, albluwi2020using, aivaloglou2016kids, techapalokul2017understanding, bai2019amelioration, delev2017static}. 
Recently, many tools for detecting code quality issues were developed and adapted for novice programmers, for example, \textit{AutoStyle}~\cite{choudhury2016scale}, \textit{WebTA}~\cite{ureel2019automated}, \textit{Refactor Tutor}~\cite{keuning2021tutoring}, and \textit{Hyperstyle}~\cite{birillo2022hyperstyle}. 
Such tools find code quality issues in a way that is adapted to the educational process --- the issues can be grouped by complexity or category, and the hints are usually written in a more comprehensible style than those of professional tools.

Together with new tools, many empirical studies emerged about the distribution of code quality issues in student submissions~\cite{keuning2017code, edwards2017investigating, albluwi2020using}.
For example, in Java, Keuning et al. study the Blackbox~\cite{brown2014blackbox} dataset of solutions to find the most prevalent issues among students~\cite{keuning2017code}. 
In addition, they studied how students attempt to correct these problems over time.
Similar research has been done for other languages, such as Scratch~\cite{techapalokul2017novice, aivaloglou2016kids, techapalokul2017understanding}, C++~\cite{bai2019amelioration}, and Python~\cite{liu2019static, molnar2020using, tigina2023analyzing}.
These kinds of studies can be useful both in the development of code quality detection tools and computer science courses.

However, code quality issues can be introduced by teachers and educators themselves within pre-written template code when creating a task. 
To the best of our knowledge, there are no studies addressing this issue, while a lot of modern MOOC platforms use pre-written task templates. 
Many studies analyse the code quality of students' submissions, however, if most students make or do not correct some kind of issue, the reason may not be their poor programming experience, but an issue in the initial task template. 

\textbf{JetBrains Academy and Hyperstyle}. JetBrains Academy~\cite{hyperskill} is a programming learning platform developed by JetBrains~\cite{jetbrains}.
Currently, the platform supports studying Java, Python, Kotlin, JavaScript, and Go.
The tasks on the platform may contain a template --- the code pre-written by the creator of the task, which should be supplemented with some new code by a student.

On the platform, the educational process is structured as follows: firstly, the student's solution is checked for correctness using traditional predefined tests. 
Then, if the solution is correct, the \textit{Hyperstyle}~\cite{birillo2022hyperstyle} tool is launched to check the quality of the code.
Therefore, the result of the \textit{Hyperstyle} tool can only be determined for \textit{correct solutions}, that is, solutions that pass all the tests.
After successfully passing the solution, the student receives a code quality grade on a four-point scale, and all detected issues (if any) are highlighted in the code editor~\cite{hypesrtyleDocs}.
An example of the \textit{Hyperstyle} user interface can be found in the supplementary materials~\cite{artifacts}.

In \textit{Hyperstyle}, code quality issues are divided into five categories.
The \textit{Code style} category contains issues with formatting, \textit{e.g.}, missed whitespaces or incorrect indentation.
The \textit{Code complexity} category considers design problems, \textit{e.g.}, a class being too complicated and not being easy to extend.
The \textit{Error-proneness} category finds potential errors and hidden bugs, \textit{e.g.}, using the \texttt{switch} statement without a default case.
The \textit{Best practices} category informs students about using non-efficient constructs and practices, \textit{e.g.}, making an abstract class without abstract methods.
The last category---\textit{Minor issues}---informs about possible typos or other issues.
The first four categories affect the final grade, and the last category, \textit{Minor issues}, merely informs the students about the problems.

The maturity of JetBrains Academy, its extensive use of templates, as well as the embedded code quality tool within it make it a perfect candidate for studying code quality issues in these templates. Moreover, our preliminary analysis of the platform's tasks lead us to numerous comments made by students where they complained that their grade was lowered for an issue not committed by them, but that was already in the pre-written code. This reiterates the importance of carrying out a study such as ours.

\section{Approach}

\begin{figure*}[h]
  \centering
  \includegraphics[width=\linewidth]{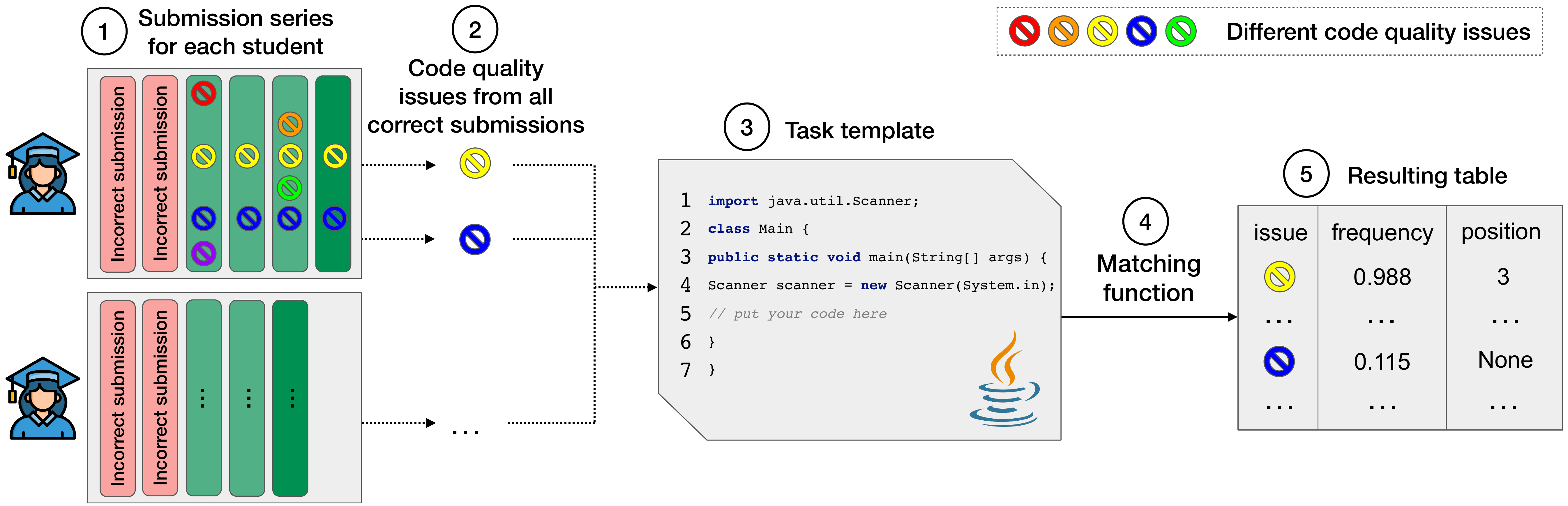}
  \caption{The general pipeline of the algorithm for detecting code quality issues in pre-written templates.}
  \label{fig:algorithm:pipeline}
  \vspace{-0.3cm}
\end{figure*}

 We have developed an algorithm to automatically detect code quality issues in pre-written task templates to help task creators detect and fix them.
Figure~\ref{fig:algorithm:pipeline} shows the general pipeline of the proposed algorithm.
Firstly, we run \textit{Hyperstyle} on solutions for problems containing a pre-written template (\circled{1} in  Figure~\ref{fig:algorithm:pipeline}), and then \circled{2} only keep the issues that are present in all of the successful attempts.
This way, we focus on issues that the students did not fix and that could thus not be introduced by them. 
Next, \circled{3} we take the original template and \circled{4} use a \textit{matching function}  to match code quality issues from the student solutions to it. 
For each issue, we have the position in the solution (line number), which allows us to extract the student's code lines and compare them with the template.
After running this matching function, \circled{5} we get the final table with the list of raw template issues.
Finally, it is possible to run a special converter to present the results in a user-friendly format.
Let us now describe this pipeline in greater detail.

\textbf{Finding candidates}. As mentioned above, code quality is only checked on \textit{correct} solutions that passed all the tests.
At this point, the student can try to fix their code quality problems or move on to the next problem.
We will refer to a sequence of successive correct solutions to a given task sent by a given user as a \textit{submission series}.

Since running the \textit{Hyperstyle} tool on the template itself does not yield correct results because of the possible false positives and compilation errors, we decided to base our approach on another premise. Our main assumption for the algorithm is that a code quality issue is located in the template if it has not been fixed throughout all the submissions in the solution series of as many students as possible.
The main reason for making this assumption is that the students know that the code in the predefined template should not be changed.

Based on this idea, for a given studied task, we find \textit{unfixed} code quality issues, \textit{i.e.}, candidates for code quality issues in pre-written templates.
Firstly, we group all submissions to each task into series by students (\circled{1} in Figure~\ref{fig:algorithm:pipeline}).
Next, we get code quality issues for each code fragment by running the \textit{Hyperstyle} tool.
Finally, \circled{2} we keep only the issues that were in each submission in the given series and remained unfixed. Such issues adhere to our assumption and constitute the initial candidates for our algorithm.

\textbf{Matching with the template}. For each detected issue that is not corrected in all solutions from the given series, we calculate its potential position in the template.
To do this, \circled{3} we first get the template for the current task, split it into lines, and \circled{4} match each of them with the corresponding line of the student code.

If a match was detected, we save the line number from the template.
If no match was found for the string, we use \textit{None} as the position, which indicates that the matching function was unable to determine the line from the pre-written template that contains this issue.
However, simply removing an issue marked with \textit{None} might not be the best solution.
This has to do with the fact that students do still change template code and move it around sometimes, so a common issue in a line that does not have an exact match with the template line can still originate from the template.
In our validation (see Section~\ref{sec:validation}), we discovered many such cases.

Finally, we calculate the frequency of appearance of each particular issue with a particular position in all series (\textit{i.e.}, the percentage of students who kept this issue unchanged in the given task).

\textbf{Filtering and labeling}. After processing the intermediate results with the matching function, \circled{5} we obtain a table (unique for each task). In this table, we display code quality issues, their position in the pre-written template, and their frequencies.
The main idea of this step is to process the obtained results to keep only true positive candidates. 
Our primary goal is to help educators and content creators to eliminate code quality issues in templates, however, it is not possible to do this fully automatically.
After the algorithm is run, the results are passed to the task creators for further review and for correcting the issues, if necessary.
Since intermediate results usually contain too many candidates, it is important for us to filter them so that the algorithm has fewer false positives.
The processing consists of two steps that we now describe in detail.

\textit{Frequency threshold}. Firstly, we apply basic filtering by the frequency of issues, \textit{i.e.}, the percentage of students who had this issue unfixed in their solution series.
If an issue remains unfixed only for a small percentage of users, it is most likely not an issue in the template and the students just did not fix it.
We chose a threshold of 10\,\%, meaning that the issue remains if at least 10\,\% of all students have not corrected it. 
This threshold was chosen empirically during our preliminary evaluation, however, it can be set independently when starting this step.
A low threshold value like this represents the initial rough filtering, aimed only at removing the least relevant results, after which the remaining candidates move on to labeling.

\textit{Labeling issues by frequency}. Secondly, while the filtering by frequency removed the least relevant cases, the remaining ones also differ in importance based on their frequency.
There are certain cases when the same issue does not get fixed by a lot of students while not originating from the template.
For example, many students forget to leave an empty line after imports.

Based on our observations, we defined three \textit{classes} of issues as follows. If the frequency of an issue is over 50\,\%, the issue class is \textit{Template} (which means that we consider it coming from the template), otherwise it is \textit{Typical} (a typical issue made by students in their solutions).
In addition, we divide \textit{Typical} issues into two more classes: \textit{Common Typical} issues (frequency from 25\,\% to 50\,\%) and \textit{Rare Typical} issues (frequency from 10\,\% to 25\,\%).

The presented thresholds were determined by us in preliminary experiments and can also be adjusted when applying the algorithm. 
Since the goal of the algorithm is to detect issues in templates, the most interesting issues are of the \textit{Template} class. 
The reason for not simply filtering out anything below this threshold is that common issues could be of interest to task creators even when they are not part of a template.
If many students repeat the same issue in their own code, this might indicate a problem in the task itself or its formulation.
Having a dedicated list of \textit{Common Typical} issues for a specific task can help task creators find the most suspicious cases, \textit{e.g.}, unfamiliar concepts or difficult places in the course.

\begin{figure*}[h]
  \centering
  \includegraphics[width=\linewidth]{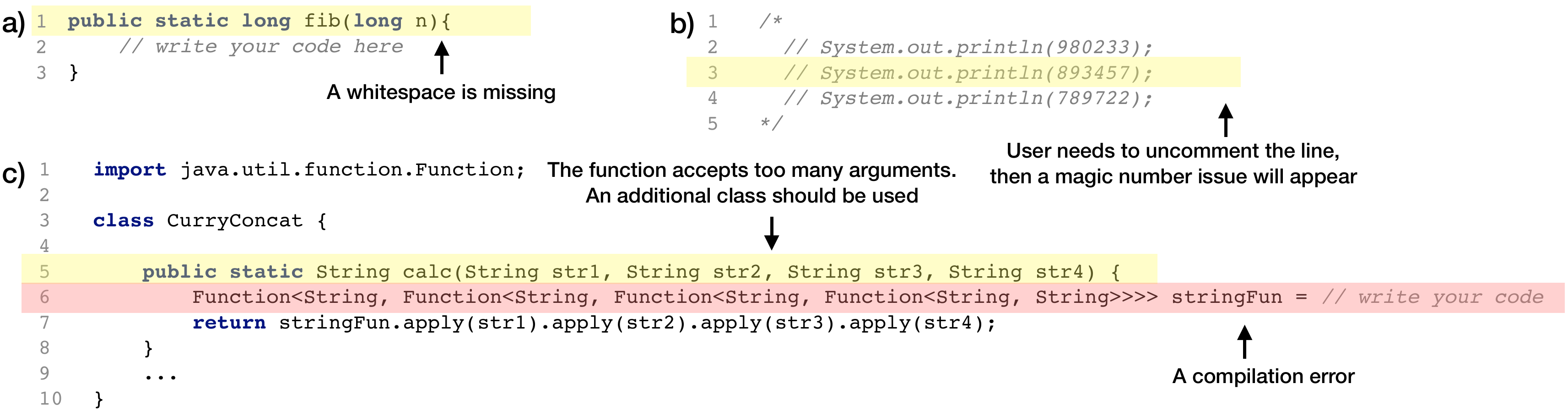}
  \caption{Several examples of the found \textit{template} issues. (a) Incorrect formatting (missing whitespace) in the pre-written template. (b) A magic number (direct usage of a number without marking it as a constant value) is hidden in the initial template, but will appear after uncommenting the line. (c) Using too many arguments in the function, the issue is hidden because the initial code fragment does not compile. }
  \vspace{-0.3cm}
  \label{fig:validation:issues:template}
\end{figure*}

\textbf{Converting to a user-friendly format}. In the end, the algorithm produces a CSV file with the following information: task ID, a unique code quality issue ID, its frequency, and the position in the template.
To provide a user-friendly output format, we implemented a converter that adds a description for each issue, a link to the task on the JetBrains Academy platform, and the newly-defined \textit{class} of the issue: \textit{Template}, \textit{Common Typical}, or \textit{Rare Typical}. 
Additionally, the user can select and save random student solutions containing a given issue in a given task, which may help them in correcting the issues.
This format was approved by the developers of the JetBrains Academy platform.

\vspace{-0.1cm}
\section{Dataset}

To evaluate the approach and conduct the analysis of code quality issues in pre-written templates, we collected all successful submissions in Java courses on the JetBrains Academy platform in a period of one year (September 1st, 2020 -- September 1st, 2021).
This resulted in \javasubmissionsinitial Java submissions of \javastudents students for \javasteps tasks, which all have pre-written templates. 
The mean length of a template is 21.6 lines, and the longest template contains as many as 314 lines. 
This indicates the importance of detecting potential code quality issues in the templates, since such an amount of code is unlikely to be without any issues.

The dataset was provided to us by the JetBrains Academy platform for this study and cannot be shared with the community because of privacy concerns. Even though it does not contain any personal information like names or emails, it contains private solutions of the users. However, the described approach itself, as well as its source code, does not rely on the platform in any way and can be used for any other dataset. The data we collected allows us to carry out the first study of this kind and provide insights into the pervasiveness of code quality issues in templates.

\vspace{-0.1cm}
\section{Validation}~\label{sec:validation}

\vspace{-0.2cm}

To check the correctness of the proposed algorithm, we conducted a manual validation of the data obtained on a subset of tasks.
The goal of the manual labeling was to check each case whether it is actually an issue in the template or not, as well as to see whether the tool missed some code quality issues.
In this section, we describe this process, its results, as well as our observations.

\textbf{Methodology}. For the manual analysis, we randomly selected \validationsteps tasks from the dataset (about 20\,\% of the entire dataset) and launched the described algorithm on it. For each studied task and each type of detected issue, five random solutions were extracted that contain it.
Next, for each task, we took its pre-written template and ran the \textit{Hyperstyle} tool independently to define the possible set of code quality issues in it. 
However, as we noted above, the template can have extra issues that will disappear after solving the task, or have no issues at all because the template code will not compile.
Therefore, for each issue in each task, the first author manually reviewed the sampled solutions of the students and the template.
The first author has more than 3 years of industrial programming experience and more than 5 years of teaching experience.

Even though the algorithm provides the final table with code quality issues in the \textit{Template} class, to find potential false negatives, the manual labeling was carried out for all the outputted issues, including the \textit{Common Typical} and \textit{Rare Typical}.

\textbf{Results and Discussion}. Overall, in the studied sample, 78 issues were detected as template-based, of which the manual review confirmed 63 as true positives, resulting in a \textbf{precision of 80.8\,\%}. As for the false negatives, 10 issues that were labelled as \textit{Common Typical} were actually from templates, and 13 issues were labelled as \textit{Rare Typical}.
Recall was calculated as a percentage of true positive \textit{Template} issues found by the algorithm among all \textit{Template} issues labeled manually.
In this experiment, we got a \textbf{recall of 73.3\,\%}.
The resulting numbers are promising and show the usefulness of the proposed approach for detecting issues in task templates. The presence of some false positives and false negatives demonstrates that the thresholds have to be tuned and studied further. 

Figure~\ref{fig:validation:issues:template} demonstrates some examples of the detected template errors. 
Sometimes, the template code contains a minor code style issue such as a missing whitespace in Figure~\ref{fig:validation:issues:template}a.
Figure~\ref{fig:validation:issues:template}b presents a more difficult case, in which a template has an issue, but it is hidden within a comment and can not be found by code quality assessment tools. 
To solve the task, a student has to uncomment the code (the task shows how comments work), and in this case, a magic number appears --- a number that is not defined as a constant value.
Figure~\ref{fig:validation:issues:template}c represents the hardest case: the initial pre-written template does not compile, and the \textit{Hyperstyle} tool will not find any issues.
However, when a student solves the task, an issue in the template will appear --- using too many arguments in a function.

As discussed earlier, sometimes the presence of issues has to do with the task itself. For example, the described \textit{Magic Number} issue can be found in many tasks, because the tasks commonly ask students to do something an arbitrary number of times. This can happen in the template or in the student's code, and both the task creators and the platform creators must decide whether it is appropriate. If the student is not expected to know certain code style features (\textit{e.g.}, the importance of naming variables and constants), then the code quality checks must be changed accordingly. 
At the same time, task creators should decide which rules they prefer to instill early on and fix all code quality issues in the templates.

Among the confirmed template issues, 23 issues had \textit{None} as their position. 
Such errors can be related to incorrect formatting, \textit{e.g.}, a missing empty line between methods in a class. In such cases, the algorithm cannot match the empty string with the template, since it was not there in the first place. This indicates the importance of keeping popular unfixed issues even if they are not directly matched to a template line, since they may still belong to the template.
\section{Analysis}

Having measured the performance of the approach, we conducted an analysis of the presence of issues in task templates of online learning platforms, taking JetBrains Academy as an example. We describe our methodology, the results, and potential implications.

\textbf{Methodology}. In our analysis, we set out to answer two research questions: 

\begin{enumerate}[start=1,label={\bfseries RQ\arabic*:}]
    \item How often do templates contain code quality issues?
    \item What are the most common code quality issues in templates?
\end{enumerate}

To answer these questions, we launched the algorithm on all 415 tasks in our dataset. For RQ1, we grouped the detected issues by task and counted how many templates had issues. For RQ2, we grouped the issues by their type (name) and looked at which were the most common. To see an even more high-level picture, we also grouped the issues by their five \textit{categories} (see Section~\ref{sec:RW}).

\textbf{Results}. \textit{RQ1: Presence of issues in the templates}. 
Firstly, of 415 tasks, 61 (14.7\,\%) have at least one issue in the template. This is a rather high number, as it indicates that a lot of tasks require some edits or tweaks. The largest number of issues in a single task is 15, indicating that there are tasks with an abnormally high number of issues that can make it harder for students to learn about code quality. Sometimes, the template contains similar pieces of code, and so the issues repeat. 

41 tasks (9.9\,\%) only have a single issue, and 20 (4.8\,\%) have at least two different kinds of issues: 11 tasks have two issues, six tasks have three issues, and two tasks have four issues. Finally, one task has as many as 5 different issue types (11 issues in total).
Unsurprisingly, this is the task with the largest template (314 lines). 
Firstly, it contains two cases of magic numbers. 
In addition, the template has formatting issues (three \textit{Empty Line Separator} issues and two \textit{Whitespace Around} issues), as well as several more serious issues: the absence of a call of the parent's constructor in the child class and the use of the same name for the field and the class method.

Although the algorithm still has room for improvement, these numbers, as well as the presence of particular templates with a lot of issues, indicate its usefulness and the necessity to conduct further research on other educational platforms.

\begin{figure}[h]
  \centering
  \includegraphics[width=\linewidth]{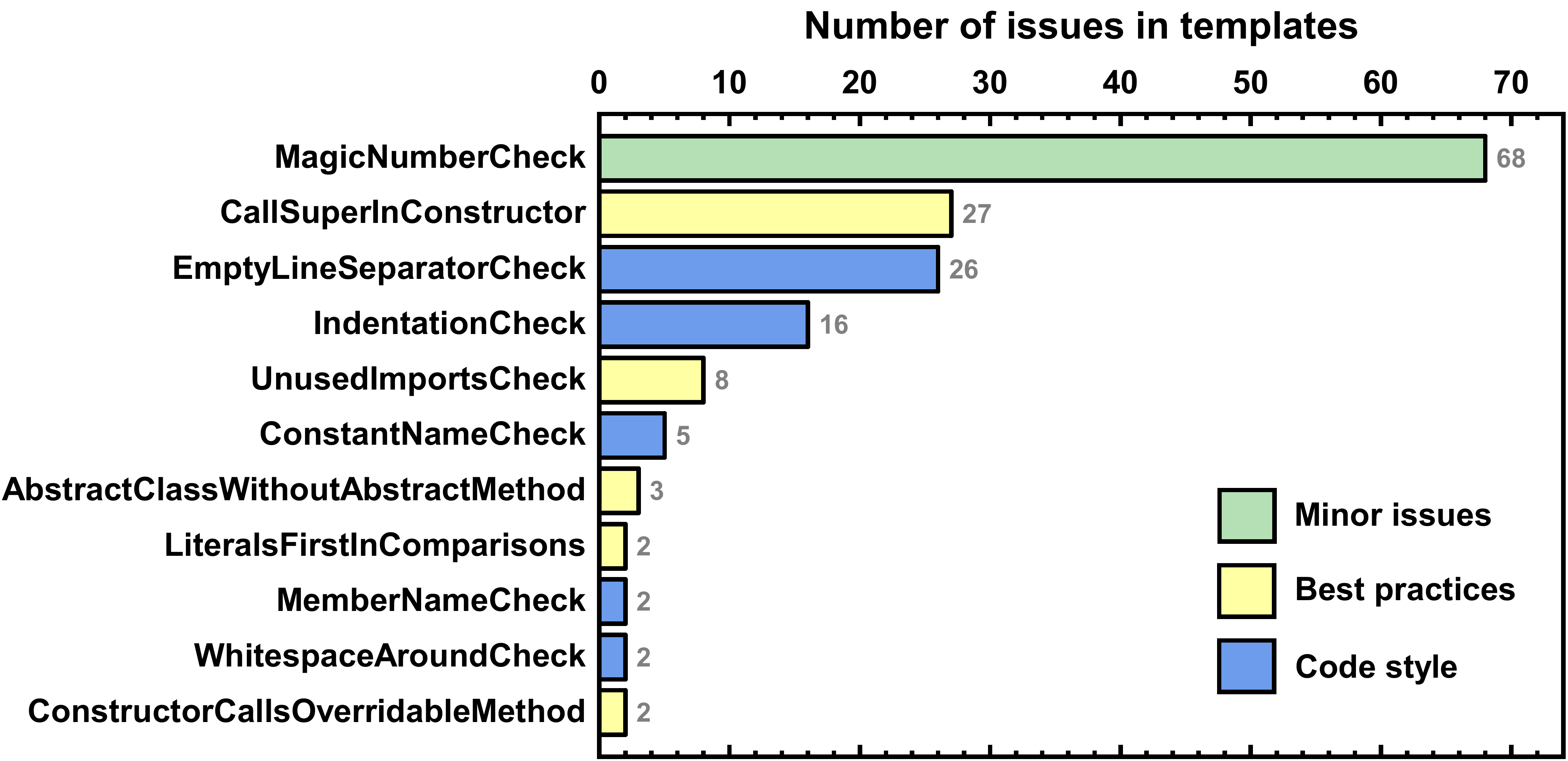}
  \caption{The number of issues of different types in the templates. The colors indicate different high-level categories.}
  \label{fig:issues}
   \vspace{-0.5cm}
\end{figure}

\textit{RQ2: Types of code quality issues in templates}.
Overall, 172 non-unique issues were detected in the templates. The distribution of those that were encountered at least two times is presented in Figure~\ref{fig:issues}. The most popular issue by far is \textit{Magic Number}, which, as discussed in Section~\ref{sec:validation}, is common both in templates and the student solutions themselves. The next most popular issue, \textit{Call Super In Constructor}, concerns omitting a call to the base class constructor when creating a child class. 
If the superclass has a constructor with no parameters or it is not accessible, then not specifying the superclass constructor to be called in the subclass constructor is a compiler error, so it must be specified.
Next, we can see formatting issues, such as \textit{Empty Line Separator} and \textit{Indentation} issues. 

When divided by categories, only three of five of them are present in the templates. The largest category, \textit{Minor issues}, contains 68 issues, however, all of them are instances of \textit{Magic Number}. The remaining issues are split approximately equally between \textit{Code Style} issues (54) and \textit{Best Practices} (50). 
Whereas \textit{Code Style} issues only affect the readability and formatting of the code, the issues from the \textit{Best Practices} category directly affect the learning of language features, and therefore, the quality of learning itself in general.

To summarize, while some formatting errors are simply actual mishaps in the template code, sometimes they appear principally from the structure of the template and the task. Code style and best practices are important to teach the students early on, and thus, the templates ideally must not contain such issues.

\section{Discussion and Implications}

\textbf{Main findings.} Our analysis showed that many programming tasks have code quality issues in their pre-written templates.
It is important to note that not all of them are merely formatting --- there are more serious problems, such as missing \texttt{super} calls in constructors, or creating abstract classes without abstract methods.
Also, there are more issues in large templates than in small ones, which makes it difficult for them to be corrected not only by the task creators but also by the students themselves.
Finally, students get frustrated if they see many issues in their solution that they did not introduce and do not receive the maximum grade because of this, as evidenced by numerous comments on the platform itself.

\textbf{Origin of issues.} One might ask --- how did these issues appear on the platform? Most platforms evolve gradually, and different features appear on them at different points in time. Only recently have the platforms started to offer such features as code quality checks, while the list of tasks is usually compiled slowly over the years. Therefore, we expect that some issues will appear --- both formatting and conceptual, since the creators of tasks might not have remembered all the code quality guidelines.

\textbf{Dataset size limitation.} In our study, we considered a relatively large dataset as an input for the algorithm, and of course, it is not always possible to collect that much data. 
However, \textit{even one solution could be enough}, since then the frequency of the issue not being corrected in such a case would be maximal (100\,\%).
In this case, not all issues might be found --- the task author may keep in mind a solution in which the quality of the code is excellent, but students may solve the task in some other way. For example, they can read numbers using \texttt{scanner.hasNext} instead of looping over \texttt{N} iterations.
The \textit{optimal} strategy for finding all issues in a template is to provide the algorithm with multiple reference solutions using different algorithms to solve the task.

\textbf{Possible usage.} This algorithm can be used in three main scenarios. (1) If a MOOC platform does not use code quality assessment tools, then before embedding them, the course creators can run the algorithm to find and eliminate code quality problems in templates. (2) If a new task needs to be added to the platform, it is enough to provide the algorithm with at least one reference solution (or several with different approaches) to find code quality issues in the template and eliminate them before publishing the task. (3) When conducting research on code quality issues in student submissions, it is possible to use the algorithm to eliminate those that were not made by students, as was done in our recent work~\cite{tigina2023analyzing}.

\textbf{Fixed issues.} To highlight the practical importance of our research, we provided the detailed output of our approach to the platform developers to fix the discovered template issues.
So far, 51 issues have been fixed in 27 tasks, with 32 of them being from the \textit{Code style} category, 14 from the \textit{Best practices} category, and 5 \textit{Minor issues}.
The platform creators found the results of the algorithm to be useful, as well as the way they are presented.
In the future, they plan to fix all the remaining issues and employ this algorithm before publishing new tasks. Moreover, they also found the \textit{Common Typical} issues to be of use as well for re-evaluating the tasks where students repeat the same mistake a lot. Overall, we believe that our algorithm and the results of this study can be useful for increasing the quality of courses in the future.
\section{Threats to Validity}

The fact that the proposed algorithm is the first of its kind, as well as the presented analysis, imposes certain threats to the validity of our work.

\textbf{Thresholds.} The thresholds used for filtering and labeling candidates were obtained by the authors of the paper after conducting preliminary experiments and manually evaluating the results. Further research is required to determine whether they are optimal. At the same time, the current validation results demonstrate adequate performance, and the feedback from the JetBrains Academy developers indicates that the output of the algorithm is useful.

\textbf{Manual validation.} While the labeling of issues during the manual validation was carried out by a single expert, they have several years of relevant experience.

\textbf{Generalizability.} The results of our analysis may not generalize to other platforms or even other languages on the same platform in terms of the distribution of issues. 
However, this study shows that even professional task creators often introduce quality issues and their content should be validated.
Finally, the goal of this analysis was to present the first general insights and highlight the importance of this type of research for further extensive investigations.

While these threats are important to highlight, we believe they do not invalidate the usefulness of our tool for further research and practical applications, as well as the importance of our pilot study for motivating this research.

\section{Conclusion and future work}

In this work, we presented the first algorithm for detecting code quality issues in the templates of tasks on educational platforms for learning programming, validated it, and conducted a first-of-its-kind analysis of the presence of such issues in templates on the JetBrains Academy platform. Since it is often not possible to detect code quality issues in the template itself, our algorithm is based on the idea of analyzing commonly unfixed issues in the submissions of multiple students. 
The manual validation conducted on 85 Java tasks on the JetBrains Academy platform (about 20\,\% of the entire dataset) demonstrated a precision of 80.8\,\% and a recall of 73.3\,\%. Our manual investigation also discovered many examples where the issues can be located only with the help of our approach --- when they are ``hidden'' in the template or move between different solutions. Finally, based on our approach, we carried out a study on 415 tasks on the same JetBrains Academy platform. Our results demonstrated that as much as 14.7\,\% of the tasks have issues in their templates, with 4.8\,\% having at least two different kinds. We also analyzed and discussed specific issues that were most prevalent, as well as examples of tasks with an abnormal number of issues.

The results of our work indicate that this research should be continued. In future work, we plan to extend the research to other languages as well as other platforms, and to compare them in terms of their distribution. 
Also, since the algorithm proved its practical use, we plan to continue using it to help with the improvements of the JetBrains Academy platform by running this algorithm and its next versions on other languages and for new tasks on the platform. We hope that our approach and our pilot study can help future researchers and practitioners with perfecting students' experience when studying programming on online platforms.

\bibliographystyle{ACM-Reference-Format}
\balance
\bibliography{sample-base}

\end{document}